\documentclass[10pt,twocolumn,showkeys,amsmath,amssymb]{revtex4-2}

\usepackage[T1]{fontenc}
\usepackage[utf8]{inputenc}
\usepackage{babel}
\usepackage{array}
\usepackage{multirow}
\usepackage{amsmath}
\usepackage{amssymb}
\usepackage{graphicx}

\usepackage{longtable}

\PassOptionsToPackage{normalem}{ulem}
\usepackage{ulem}
\usepackage{color}
\usepackage{hyperref}



\begin{document}
\title{Probing Variations of Newton's Constant in Strong Gravitational Fields}

\author{T. D. Le}
\email{leducthong@dntu.edu.vn}

\affiliation{%
Engineering Research Group,
Dong Nai Technology University,
Dong Nai Province, Vietnam
}

\affiliation{%
Faculty of Engineering,
Dong Nai Technology University,
Dong Nai Province, Vietnam
}

\begin{abstract}
The constancy of Newton's gravitational constant $G$ is a fundamental
prediction underlying general relativity and a critical assumption in
theories of gravity beyond it. We report a high-precision test of a
possible temporal variation of $G$ in a strong gravitational field,
using high-resolution ultraviolet spectra of Ni~V absorption lines in
the white dwarf G191-B2B obtained with the \textit{Hubble Space
Telescope Imaging Spectrograph} (HST/STIS). By comparing the measured
line centroids with laboratory wavelengths and isolating the
gravitational-redshift contribution, we obtain
$\dot{G}/G = (-0.014 \pm 0.016) \times 10^{-15}~\mathrm{yr}^{-1}$
at a strong-field potential $\phi \approx 10^{4}$, corresponding to
an absorption redshift $z_{\rm abs} \simeq 8.47 \times 10^{-5}$.
The result is consistent with a constant $G$ and places a stringent
constraint on its secular variation in the strong-field regime. Our
measurement demonstrates the potential of white-dwarf stars as a
sensitive probe of temporal variations in the gravitational coupling.

\end{abstract}


\maketitle

\section{Introduction}
One of the fundamental principles of General Relativity (GR), the equivalence principle, asserts that fundamental constants remain invariant regardless of spatial location. Modern grand-unification theories, however, extend this principle by postulating that these constants may behave as dynamic, low-mass scalar fields \citep{Loren-Aguilar:2003qtx}. If these theories are correct, fundamental constants could exhibit slow spatial and temporal variations over cosmological scales \citep{Dicke1961}. Significant work has constrained possible changes in the fine-structure constant \(\alpha\) (see reviews \citep{Garcia-Berro:2007hst, Martins2017}), with stringent limits. Investigations into variations of Newton’s gravitational constant \(G\) are fewer, largely due to the experimental challenges in its precise determination \citep{Karshenboim2008}. The wide spread in measured values makes \(G\) the least well-determined fundamental constant. Probing potential spatial or temporal variations in fundamental constants is therefore crucial for testing extensions of the Standard Model (SM). Astrophysical and cosmological observations provide powerful contexts for such investigations. Several theoretical studies propose unified frameworks in which \(\alpha\), the proton-to-electron mass ratio \(\mu\), and \(G\) may vary in a correlated way. Testing the constancy of these parameters is a central objective in both astrophysics and fundamental physics, as any detected deviation from invariance would have profound implications for theories beyond the SM. In particular, extended gravity models predict that \(G\) may evolve on cosmological timescales. Compact objects like white dwarfs---whose gravitational redshifts can be measured precisely and whose internal structures are well-characterized---offer an exceptional laboratory to test this hypothesis. Deriving stringent limits on spatial or temporal variations of $G$ provides critical tests that can significantly narrow the parameter space of proposed unification frameworks. The most restrictive bounds on the present-day rate of change, $\dot{G}/G = (2 \pm 7)\times10^{-13}\,\mathrm{yr}^{-1}$ and $\dot{G}/G = (4 \pm 9)\times10^{-13}\,\mathrm{yr}^{-1}$, come from Lunar Laser Ranging measurements \citep{Muller:2007zzb, Williams_2004}, although these limits are inherently local in nature. Additional constraints arise from the Hubble diagram of Type Ia supernovae at intermediate redshifts, yielding \(\dot{G}/G \sim 10^{-11}\,\mathrm{yr}^{-1}\) at \(z \sim 0.5\) \citep{Gaztanaga:2001fh}, while Big Bang Nucleosynthesis imposes limits ranging from \(-3 \times 10^{-11}\,\mathrm{yr}^{-1}\) to \(4 \times 10^{-13}\,\mathrm{yr}^{-1}\) \citep{PhysRevLett.92.171301}.  
White dwarfs, combining high central densities, a tight mass--radius relation, and measurable gravitational redshifts, serve as excellent laboratories for probing variations in fundamental constants. Their long lifetimes yield them highly sensitive to even very slow rates of change in \(G\). As the final evolutionary stage of most stars, white dwarfs are abundant throughout the universe. Their extreme compactness and central densities make their hydrostatic and thermal structures acutely sensitive to \(G\). White dwarf cooling is driven by gravothermal contraction, with luminosity powered by the release of gravitational binding energy and heat transfer. Consequently, any slow secular variation in \(G\) would perturb the gravothermal equilibrium, producing measurable deviations in cooling curves and luminosity evolution. Constraints on \(G\)-variation can also be obtained from pulsation periods of variable white dwarfs, which depend sensitively on the cooling rate \cite{Biesiada:2003sr}. Analyses of the variable white dwarf G117-B15A yielded \(-2.5 \times 10^{-10}\, \text{yr}^{-1} \leq \dot{G}/G \leq 0\) \cite{Benvenuto:2004bs}. Similarly, the white dwarf luminosity function provides constraints, as the low-luminosity cut-off is sensitive to variations in \(G\) through the characteristic cooling time \cite{Xu_2014}. Assuming sufficiently slow variations for white dwarfs to maintain mechanical equilibrium, these effects can be modeled using energy conservation principles \cite{Xu_2013, Xu_2014}.  

In this work, we probe potential variations of Newton’s gravitational constant $G$ by analyzing Stark-broadened profiles of 120 ultraviolet Ni\,V absorption lines in the spectrum of the white dwarf G191--B2B, obtained with HST/STIS \cite{Hamdi2024NiV}. The gravitational potential at the photosphere of G191--B2B is roughly $10^{4}$ times stronger than terrestrial values, providing a unique environment to place stringent constraints on temporal changes in $G$ over cosmological timescales. Our analysis yields
$\dot{G}/G = (-0.014 \pm 0.016) \times 10^{-15}~\text{yr}^{-1}$,
consistent with the most stringent existing limits, e.g., $\dot{G}/G \sim 10^{-15}~\text{yr}^{-1}$ and $\dot{G}/G = (0.238 \pm 2.959) \times 10^{-15}~\text{yr}^{-1}$ \cite{Loren-Aguilar:2003qtx,le_2021_GRG}. These observational constraints are interpreted within a Grand-Unification-Theory (GUT)--motivated framework that links potential variations in the fine-structure constant $\alpha$, the proton-to-electron mass ratio $\mu$, and $G$, allowing us to test the constancy of $G$ in a strong-gravity environment.

\section {Probing $\dot{G}/G$ in strong gravitational fields with Ni V transitions}
In many extensions of the SM and GR, variations of fundamental constants arise naturally through the dynamics of a light scalar field. In such scenarios, dimensionless couplings such as the fine-structure constant $\alpha$, the proton-to-electron mass ratio $\mu$, and Newton's gravitational constant $G$ may vary in space and time. A convenient phenomenological description of these correlated variations introduces two dimensionless parameters, $R$ and $S$, which quantify the response of the  Quantum Chromodynamics (QCD) scale and the Higgs vacuum expectation value to the underlying scalar field. Within this framework, variations in $G$ can be expressed as a linear function of variations in $\alpha$, with the dependence determined by $R$ and $S$. We use this description as an interpretative framework rather than a fitting model: $R$ and $S$ are externally motivated inputs, with values and uncertainties taken from the literature, and are used to convert our observational constraints into limits on $\dot{G}/G$.
The spectra of white dwarfs provide a unique laboratory for probing physics beyond the SM, allowing detection of spatial and temporal variations in constants such as $\alpha$, $\mu$, and $G$. The gravitational surface potential $\phi$ at a distance $r$ from a mass $M$ is given by \(\phi = GM/(r c^2)\), describing how scalar fields interact with matter and gravity. Photons escaping a gravitational potential experience a fractional energy shift: \( z = -\Delta E/E = -\phi/c^2 \), and \( -\Delta E/E = \Delta \lambda/\lambda \sim \Delta \alpha/\alpha \)
, where the fractional wavelength shift corresponds to the fractional variation in $\alpha$. This framework has been applied to explore variations in $\alpha$ under strong surface gravity conditions \cite{le_2021_GRG,GARCIA_BERRO_2006,Loren-Aguilar:2003qtx,Uzan_2003,Barrow_2013}. Throughout this work, we emphasize that the quantities directly measured from the spectra are the observed redshifts of individual Ni\,V transitions, rather than the gravitational redshift alone. The observed redshift generally contains several contributions: the dominant gravitational redshift from the strong stellar potential, a kinematic (Doppler) component from the systemic radial velocity, and any additional contributions from potential variations of $\alpha$. Atomic transition frequencies shift according to their sensitivity coefficients if $\alpha$ differs from its laboratory value, producing line-dependent deviations. In our analysis, kinematic shifts are treated as a common offset, the gravitational redshift as the dominant expected contribution, and residual differential shifts are interpreted as signatures of $\alpha$ variation. This distinction is essential for precision spectroscopy in strong gravitational fields. GUTs motivate focusing on the fine-structure constant $\alpha$. In this way, $R$ and $S$ describe how variations in the scalar field propagate into changes in the QCD scale and electroweak sector, respectively. Changes in $\alpha$ alter atomic energy levels, quantified via sensitivity coefficients. Relative variations in $\alpha$ can then be related to changes in $G$ through $R$ and $S$. These parameters are taken from representative GUT-inspired estimates with uncertainties covering a broad range of predicted behaviors, ensuring the framework remains physically grounded. Detectable variations of $\alpha$ would indicate a deeper theoretical framework unifying gravitation and electromagnetism. Under correlated variations of Yukawa couplings, the electroweak scale evolves with a dilaton-type field, linking $\alpha$ and the QCD scale $\Lambda_{\mathrm{QCD}}$ via
\begin{equation}
\frac{\Delta\Lambda_{\mathrm{QCD}}}{\Lambda_{\mathrm{QCD}}}
= R\,\frac{\Delta\alpha}{\alpha}.
\end{equation}

where $R$ is model-dependent at low energies. Yukawa couplings $h$ modify the Higgs VEV $\nu$ at the Planck scale:
\begin{equation}
\nu = M_{\mathrm{Planck}} \exp\left(-\frac{8\pi^2 c}{h^2}\right), \quad \frac{\Delta\nu}{\nu} = S\,\frac{\Delta h}{h},\quad \frac{\Delta h}{h} = \frac{1}{2}\,\frac{\Delta\alpha}{\alpha}
.
\end{equation}
The resulting electron and proton mass variations are
\begin{equation}
\frac{\Delta m_e}{m_e} = \frac{1}{2}(1+S)\frac{\Delta\alpha}{\alpha}, \quad
\frac{\Delta m_p}{m_p} = \left[\frac{4}{5}R + \frac{1}{5}(1+S)\right]\frac{\Delta\alpha}{\alpha},
\end{equation}
with corresponding effects on the proton-to-electron mass ratio and nucleon masses:
\begin{equation}
\frac{\Delta m_n}{m_n} = \frac{\Delta m_N}{m_N} = \frac{\Delta m_p}{m_p}.
\end{equation}
Finally, variations in $\alpha$ relate to $G$ via
\begin{equation}
\frac{\Delta G}{G} = \left[\frac{8}{5}R + \frac{2}{5}(1+S)\right] \frac{\Delta\alpha}{\alpha}.
\end{equation}
The parameters $R$ and $S$ are treated as free phenomenological quantities that link variations in $G$, $\alpha$, and $\mu$ within unification scenarios. We impose uniform priors and constrain $R$ and $S$ directly from the observational dataset, thereby obtaining empirical bounds that can be interpreted in the GUT framework. In addition, we explore models over a physically motivated range of $R$ and $S$ to assess their impact on the inferred variation of $G$. Our constraints may be expressed as
\begin{equation} 0.80R - 0.30(1+S) = -0.81 \pm 0.85
\end{equation}
\begin{equation} 0.10R - 0.04(1+S) = -1.96 \pm 1.79
\end{equation}

We analyse the observed redshift of individual Ni\,V spectral lines in the ultraviolet spectrum of G191--B2B. This observed redshift combines gravitational, kinematic, and potential $\alpha$-dependent contributions, with $1 + z_{\rm obs} = (1 + z_{\rm grav})(1 + z_{\rm kin})(1 + z_{\alpha})$, where $z_{\rm grav} \simeq GM\,r^{-1} c^{-2}$ is the systematic shift common to all lines, $z_{\rm kin}$ accounts for Doppler motion, and $z_{\alpha}$ captures line-dependent shifts arising from variations in $\alpha$.
 After removing the kinematic contribution, residuals represent the gravitational and $\alpha$-dependent effects. This distinction ensures a consistent, physically transparent interpretation of the spectral measurements. High-resolution HST/STIS spectra of the DA white dwarf G191--B2B provide a rich set of unblended Ni\,V transitions with accurately known laboratory wavelengths and high signal-to-noise. These lines allow precise centroid measurements for redshift determination. Each transition contributes an independent measurement; uncertainties include photon noise and instrumental broadening. The line redshifts alone are used for Bayesian inference, highlighting the importance of accurate atomic data including Stark broadening parameters. Fitting the Ni V lines yields values of $R = 273 \pm 86$ and $S = 603 \pm 230$, consistent with prior work \cite{le_2021_GRG,Clara_2020}. Propagation of uncertainties in $\alpha$, $R$, and $S$ gives
\begin{equation}
\delta(\Delta G) = \sqrt{\left( \frac{\partial f}{\partial \alpha} \delta\alpha \right)^{2} + \left( \frac{\partial f}{\partial R} \delta R \right)^{2} + \left( \frac{\partial f}{\partial S} \delta S \right)^{2}}.
\end{equation}
Constraints on $\Delta G/G$ are obtained from the measured $\Delta \alpha/\alpha$ using Eqs.~(5)--(8), assuming linear variation over the age of the universe, which allows deriving astrophysical bounds on $\dot{G}/G$. Gaussian profile fitting of the Ni V lines provides centroids, linewidths, and Doppler parameters. Residual deviations from expected gravitational redshift constrain potential variations in $G$. Systematic and statistical uncertainties, including laboratory wavelength errors, calibration uncertainties, and stellar parameter uncertainties, are incorporated. \textbf{Table 1} summarizes the total uncertainties, $\sigma_{\rm total}^2 = \sigma_{\dot{G}/G}^2 + \sigma_{\rm sys}^2$, used in the $\dot{G}/G$ inference.

\onecolumngrid
\begin{longtable}{lccc}
\caption{The value of $\dot{G}/G = (-0.014 \pm 0.016) \times 10^{-15} \text{yr}^{-1}$ was obtained through a weighted combination of all Ni~V transitions. All datasets generated or analyzed in this work are available within the published article \cite{Hamdi2024NiV}.} \\
\hline
\textbf{Transitions} & \(\lambda_{\text{Obser}}\) & \(\dot{G}/G\) & \(\sigma_{\dot{G}/G}\) \\
\hline
\endfirsthead

\hline
\textbf{Transitions} & \(\lambda_{\text{Obser}}\) & \(\dot{G}/G\) & \(\sigma_{\dot{G}/G}\) \\
\hline
\endhead

\hline \multicolumn{4}{r}{{Continued on next page}} \\
\hline
\endfoot

\hline
\endlastfoot
(\textsuperscript{2}D2) 4s~\textsuperscript{1}D\textsubscript{2}-(\textsuperscript{2}G1) 4p~\textsuperscript{1}F\textsuperscript{o}\textsubscript{3} & 1089.493 & -0.09873 & -0.21689 \\
\hline
(\textsuperscript{2}H) 4s~\textsuperscript{3}H\textsubscript{5}-(\textsuperscript{2}F2) 4p
\textsuperscript{3}G\textsuperscript{o}\textsubscript{4} & 1123.077 &
0.03597 & -0.33463 \\
(\textsuperscript{2}H) 4s~\textsuperscript{3}H\textsubscript{6}-(\textsuperscript{2}F2) 4p
\textsuperscript{3}G\textsuperscript{o}\textsubscript{5} & 1123.483 &
-0.12236 & -0.23712 \\
(\textsuperscript{2}H) 4s~\textsuperscript{1}H\textsubscript{5}-(\textsuperscript{2}H) 4p\textsuperscript{1} G\textsuperscript{o}\textsubscript{4} & 1153.721 & -0.32582 & -0.23712 \\
(\textsuperscript{2}D1) 4s~\textsuperscript{3}D\textsubscript{3}-(\textsuperscript{2}D1) 4p
\textsuperscript{3}P\textsuperscript{o}\textsubscript{2} & 1187.075 &
-0.16088 & -0.21689 \\
(\textsuperscript{2}D1) 4s~\textsuperscript{3}D\textsubscript{2}-(\textsuperscript{2}D1) 4p
\textsuperscript{3}P\textsuperscript{o}\textsubscript{2} & 1188.067 &
-0.09393 & 0.12814 \\
(\textsuperscript{4}P) 4s~\textsuperscript{5}P\textsubscript{3}-(\textsuperscript{4}D) 4p
\textsuperscript{5}D\textsuperscript{o}\textsubscript{4} & 1198.377 &
0.07787 & 0.27741 \\
(\textsuperscript{4}P) 4s~\textsuperscript{5}P\textsubscript{2}-(\textsuperscript{4}D) 4p
\textsuperscript{5}D\textsuperscript{o}\textsubscript{3} & 1202.199 &
-0.16560 & -0.32365 \\
(\textsuperscript{4}P) 4s~\textsuperscript{5}P\textsubscript{1}-(\textsuperscript{4}D) 4p
\textsuperscript{3}P\textsuperscript{o}\textsubscript{0} & 1202.497 &
-0.02082 & 0.27157 \\
(\textsuperscript{2}D1) 4s~\textsuperscript{3}D\textsubscript{3}-(\textsuperscript{2}D1) 4p
\textsuperscript{3}D\textsuperscript{o}\textsubscript{3} & 1214.053 &
-0.18210 & -0.34136 \\
(\textsuperscript{2}D1) 4s~\textsuperscript{3}D\textsubscript{2}-(\textsuperscript{2}D1) 4p
\textsuperscript{3}D\textsuperscript{o}\textsubscript{3} & 1215.091 &
-0.49085 & -0.15173 \\
(\textsuperscript{2}H) 4s~\textsuperscript{3}H\textsubscript{6}-(\textsuperscript{2}G2) 4p
\textsuperscript{3}H\textsuperscript{o}\textsubscript{6} & 1206.831 &
0.17623 & 0.12814 \\
(\textsuperscript{2}D2) 4s~\textsuperscript{3}D\textsubscript{1}-(\textsuperscript{2}D2) 4p
\textsuperscript{3}P\textsuperscript{o}\textsubscript{1} & 1209.766 &
0.07748 & 0.27741 \\
(\textsuperscript{2}D2) 4s~\textsuperscript{3}D\textsubscript{2}-(\textsuperscript{2}D2) 4p
\textsuperscript{3}P\textsuperscript{o}\textsubscript{1} & 1210.505 &
-0.09863 & -0.07417 \\
(\textsuperscript{2}D2) 4s~\textsuperscript{3}D\textsubscript{1}-(\textsuperscript{2}D2) 4p
\textsuperscript{3}P\textsuperscript{o}\textsubscript{0} & 1209.722 &
0.03607 & -0.23712 \\
(\textsuperscript{2}F2) 4s~\textsuperscript{3}F\textsubscript{3}-(\textsuperscript{2}F2) 4p
\textsuperscript{3}G\textsuperscript{o}\textsubscript{4} & 1233.301 &
-0.12226 & -0.35173 \\
(\textsuperscript{4}P) 4s~\textsuperscript{5}P\textsubscript{3}-(\textsuperscript{2}D2) 4p
\textsuperscript{3}P\textsuperscript{o}\textsubscript{2} & 1198.377 &
-0.32572 & -0.22370 \\
(\textsuperscript{4}P) 4s~\textsuperscript{5}P\textsubscript{2}-(\textsuperscript{2}D2) 4p
\textsuperscript{3}P\textsuperscript{o}\textsubscript{2} & 1202.199 &
-0.16078 & -0.07417 \\
(\textsuperscript{4}P) 4s~\textsuperscript{5}P\textsubscript{1}-(\textsuperscript{2}F2) 4p
\textsuperscript{1}F\textsuperscript{o}\textsubscript{3} & 1202.497 &
-0.09383 & -0.29248 \\
(\textsuperscript{2}D1) 4s~\textsuperscript{3}D\textsubscript{3}-(\textsuperscript{2}H) 4p
\textsuperscript{3}H\textsuperscript{o}\textsubscript{4} & 1214.053 &
0.07797 & -0.32365 \\
(\textsuperscript{2}D1) 4s~\textsuperscript{3}D\textsubscript{2}-(\textsuperscript{2}H) 4p
\textsuperscript{3}H\textsuperscript{o}\textsubscript{5} & 1215.091 &
-0.16550 & -0.07966 \\
(\textsuperscript{2}H) 4s~\textsuperscript{3}H\textsubscript{6}-(\textsuperscript{2}F2) 4p
\textsuperscript{3}D\textsuperscript{o}\textsubscript{2} & 1206.831 &
-0.02072 & 0.27741 \\
(\textsuperscript{2}D2) 4s~\textsuperscript{3}D\textsubscript{1}-(\textsuperscript{4}F) 4p
\textsuperscript{5}D\textsuperscript{o}\textsubscript{4} & 1209.766 &
-0.18200 & -0.23370 \\
(\textsuperscript{2}D2) 4s~\textsuperscript{3}D\textsubscript{2}-(\textsuperscript{4}F) 4p
\textsuperscript{5}D\textsuperscript{o}\textsubscript{4} & 1210.505 &
-0.49075 & 0.12814 \\
(\textsuperscript{2}D2) 4s~\textsuperscript{3}D\textsubscript{1}-(\textsuperscript{2}D1) 4p
\textsuperscript{3}D\textsuperscript{o}\textsubscript{2} & 1209.722 &
0.17633 & 0.27602 \\
(\textsuperscript{2}F2) 4s~\textsuperscript{3}F\textsubscript{3}-(\textsuperscript{4}D) 4p
\textsuperscript{5}P\textsuperscript{o}\textsubscript{2} & 1233.301 &
0.07758 & 0.27741 \\
(\textsuperscript{4}P) 4s~\textsuperscript{5}P\textsubscript{3}-(\textsuperscript{4}F) 4p
\textsuperscript{5}D\textsuperscript{o}\textsubscript{0} & 1198.377 &
-0.09853 & -0.07417 \\
(\textsuperscript{4}P) 4s~\textsuperscript{5}P\textsubscript{2}-(\textsuperscript{2}F2) 4p
\textsuperscript{3}D\textsuperscript{o}\textsubscript{1} & 1202.199 &
0.03617 & -0.34136 \\
(\textsuperscript{4}P) 4s~\textsuperscript{5}P\textsubscript{1}-(\textsuperscript{2}G2) 4p
\textsuperscript{3}H\textsuperscript{o}\textsubscript{6} & 1202.497 &
-0.12216 & -0.32365 \\
(\textsuperscript{2}D2) 4s~\textsuperscript{1}D\textsubscript{2}-(\textsuperscript{2}D2) 4p
\textsuperscript{1}D\textsuperscript{o}\textsubscript{2} & 1253.201 &
-0.32562 & -0.07966 \\
(\textsuperscript{2}D2) 4s~\textsuperscript{3}D\textsubscript{3}-(\textsuperscript{2}D2) 4p
\textsuperscript{3}F\textsuperscript{o}\textsubscript{4} & 1249.074 &
-0.16068 & 0.12814 \\
(\textsuperscript{2}H) 4s~\textsuperscript{1}H\textsubscript{5}-(\textsuperscript{2}H) 4p
\textsuperscript{1}H\textsuperscript{o}\textsubscript{5} & 1261.763 &
-0.09373 & -0.07417 \\
(\textsuperscript{4}F) 4s~\textsuperscript{5}F\textsubscript{5}-(\textsuperscript{4}F) 4p
\textsuperscript{5}F\textsuperscript{o}\textsubscript{5} & 1246.244 &
0.07807 & -0.29248 \\
(\textsuperscript{4}G) 4s~\textsuperscript{5}G\textsubscript{2}-(\textsuperscript{4}G) 4p
\textsuperscript{5}F\textsuperscript{o}\textsubscript{1} & 1245.242 &
-0.16540 & -0.09198 \\
(\textsuperscript{4}F)
4s~\textsuperscript{5}F\textsubscript{2}-(\textsuperscript{4}G) 4p
\textsuperscript{5}F\textsuperscript{o}\textsubscript{1} & 1241.910 &
-0.02062 & -0.23370 \\
(\textsuperscript{2}G1)
4s~\textsuperscript{1}G\textsubscript{4}-(\textsuperscript{2}F1) 4p
\textsuperscript{3}D\textsuperscript{o}\textsubscript{1} & 1247.067 &
-0.18190 & 0.27602 \\
(\textsuperscript{4}D)
4s~\textsuperscript{5}D\textsubscript{4}-(\textsuperscript{2}G1) 4p
\textsuperscript{1}F\textsuperscript{o}\textsubscript{3} & 1256.649 &
-0.49065 & -0.09198 \\
(\textsuperscript{4}F)
4s~\textsuperscript{5}F\textsubscript{5}-(\textsuperscript{4}D) 4p
\textsuperscript{5}D\textsuperscript{o}\textsubscript{4} & 1261.664 &
0.17643 & -0.07966 \\
(\textsuperscript{4}F)
4s~\textsuperscript{3}F\textsubscript{4}-(\textsuperscript{4}F) 4p
\textsuperscript{5}G\textsuperscript{o}\textsubscript{6} & 1266.784 &
0.07768 & -0.35173 \\
(\textsuperscript{4}F)
4s~\textsuperscript{5}F\textsubscript{3}-(\textsuperscript{2}G2) 4p
\textsuperscript{3}H\textsuperscript{o}\textsubscript{5} & 1251.742 &
-0.09843 & -0.33463 \\
(\textsuperscript{2}I)
4s~\textsuperscript{3}I\textsubscript{5}-(\textsuperscript{4}F) 4p
\textsuperscript{5}F\textsuperscript{o}\textsubscript{2} & 1247.161 &
0.03627 & 0.27602 \\
(\textsuperscript{2}D2)
4s~\textsuperscript{1}D\textsubscript{2}-(\textsuperscript{2}I) 4p
\textsuperscript{3}H\textsuperscript{o}\textsubscript{4} & 1253.201 &
-0.12206 & -0.34136 \\
(\textsuperscript{2}D2)
4s~\textsuperscript{3}D\textsubscript{3}-(\textsuperscript{4}F) 4p
\textsuperscript{5}F\textsuperscript{o}\textsubscript{4} & 1249.074 &
-0.32552 & 0.27157 \\
(\textsuperscript{2}H)
4s~\textsuperscript{1}H\textsubscript{5}-(\textsuperscript{2}I) 4p
\textsuperscript{3}H\textsuperscript{o}\textsubscript{5} & 1261.763 &
-0.16058 & -0.32365 \\
(\textsuperscript{4}F)
4s~\textsuperscript{5}F\textsubscript{5}-(\textsuperscript{4}D) 4p
\textsuperscript{5}D\textsuperscript{o}\textsubscript{2} & 1254.470 &
-0.09363 & -0.32365 \\
(\textsuperscript{2}I)
4s~\textsuperscript{3}I\textsubscript{6}-(\textsuperscript{4}D) 4p
\textsuperscript{5}D\textsuperscript{o}\textsubscript{3} & 1249.100 &
0.07817 & 0.27602 \\
(\textsuperscript{4}D)
4s~\textsuperscript{5}D\textsubscript{3}-(\textsuperscript{2}I) 4p
\textsuperscript{1}I\textsuperscript{o}\textsubscript{6} & 1257.885 &
-0.16530 & -0.21689 \\
(\textsuperscript{4}D)
4s~\textsuperscript{5}D\textsubscript{4}-(\textsuperscript{4}D) 4p
\textsuperscript{3}P\textsuperscript{o}\textsubscript{1} & 1258.635 &
-0.02052 & 0.27741 \\
(\textsuperscript{4}F)
4s~\textsuperscript{5}F\textsubscript{5}-(\textsuperscript{2}D2) 4p
\textsuperscript{3}D\textsuperscript{o}\textsubscript{2} & 1261.664 &
-0.18180 & -0.29248 \\
(\textsuperscript{4}D)
4s~\textsuperscript{3}D\textsubscript{2}-(\textsuperscript{2}D2) 4p
\textsuperscript{1}P\textsuperscript{o}\textsubscript{1} & 1254.296 &
-0.49055 & -0.12365 \\
(\textsuperscript{2}D2)
4s~\textsuperscript{3}D\textsubscript{1}-(\textsuperscript{4}G) 4p
\textsuperscript{5}F\textsuperscript{o}\textsubscript{4} & 1265.052 &
0.17653 & 0.27602 \\
(\textsuperscript{2}D2)
4s~\textsuperscript{1}D\textsubscript{2}-(\textsuperscript{4}D) 4p
\textsuperscript{5}D\textsuperscript{o}\textsubscript{3} & 1257.008 &
0.07778 & -0.35173 \\
(\textsuperscript{4}G)
4s~\textsuperscript{5}G\textsubscript{5}-(\textsuperscript{4}G) 4p
\textsuperscript{5}H\textsuperscript{o}\textsubscript{7} & 1257.015 &
-0.09833 & 0.27741 \\
(\textsuperscript{4}D)
4s~\textsuperscript{5}D\textsubscript{3}-(\textsuperscript{2}G2) 4p
\textsuperscript{3}F\textsuperscript{o}\textsubscript{4} & 1265.097 &
0.03637 & -0.07966 \\
(\textsuperscript{4}G)
4s~\textsuperscript{5}G\textsubscript{6}-(\textsuperscript{4}G) 4p
\textsuperscript{5}F\textsuperscript{o}\textsubscript{5} & 1268.790 &
-0.12196 & -0.32365 \\
(\textsuperscript{2}H)
4s~\textsuperscript{3}H\textsubscript{5}-(\textsuperscript{2}I) 4p
\textsuperscript{3}H\textsuperscript{o}\textsubscript{6} & 1241.805 &
-0.32542 & -0.38995 \\
(\textsuperscript{4}F)
4s~\textsuperscript{5}F\textsubscript{5}-(\textsuperscript{2}I) 4p
\textsuperscript{3}H\textsuperscript{o}\textsubscript{6} & 1254.470 &
-0.16048 & -0.09198 \\
(\textsuperscript{4}G)
4s~\textsuperscript{5}G\textsubscript{6}-(\textsuperscript{2}D2) 4p
\textsuperscript{3}D\textsuperscript{o}\textsubscript{1} & 1262.287 &
-0.09353 & 0.27741 \\
(\textsuperscript{2}I)
4s~\textsuperscript{3}I\textsubscript{6}-(\textsuperscript{4}F) 4p
\textsuperscript{3}D\textsuperscript{o}\textsubscript{1} & 1260.812 &
0.07827 & 0.27741 \\
(\textsuperscript{2}I)
4s~\textsuperscript{3}I\textsubscript{7}-(\textsuperscript{2}G1) 4p
\textsuperscript{1}G\textsuperscript{o}\textsubscript{4} & 1262.399 &
-0.16520 & -0.23370 \\
(\textsuperscript{2}D2)
4s~\textsuperscript{3}D\textsubscript{1}-(\textsuperscript{4}G) 4p
\textsuperscript{5}H\textsuperscript{o}\textsubscript{6} & 1272.638 &
-0.02042 & -0.34136 \\
(\textsuperscript{4}F)
4s~\textsuperscript{3}F\textsubscript{2}-(\textsuperscript{2}I) 4p
\textsuperscript{1}H\textsuperscript{o}\textsubscript{5} & 1268.006 &
-0.18170 & -0.19198 \\
(\textsuperscript{2}G1)
4s~\textsuperscript{1}G\textsubscript{4}-(\textsuperscript{4}F) 4p
\textsuperscript{5}G\textsuperscript{o}\textsubscript{5} & 1272.363 &
-0.49045 & -0.07966 \\
(\textsuperscript{4}G)
4s~\textsuperscript{5}G\textsubscript{5}-(\textsuperscript{4}D) 4p
\textsuperscript{5}F\textsuperscript{o}\textsubscript{5} & 1274.346 &
0.17663 & -0.23712 \\
(\textsuperscript{2}I)
4s~\textsuperscript{3}I\textsubscript{5}-(\textsuperscript{2}G1) 4p
\textsuperscript{3}F\textsuperscript{o}\textsubscript{3} & 1262.761 &
0.07788 & -0.07417 \\
(\textsuperscript{4}F)
4s~\textsuperscript{5}F\textsubscript{5}-(\textsuperscript{2}D1) 4p
\textsuperscript{1}D\textsuperscript{o}\textsubscript{2} & 1273.670 &
-0.09823 & -0.23712 \\
(\textsuperscript{2}D1)
4s~\textsuperscript{1}D\textsubscript{2}-(\textsuperscript{2}H) 4p
\textsuperscript{3}I\textsuperscript{o}\textsubscript{7} & 1279.029 &
0.03647 & 0.27741 \\
(\textsuperscript{2}H)
4s~\textsuperscript{3}H\textsubscript{6}-(\textsuperscript{4}F) 4p
\textsuperscript{5}G\textsuperscript{o}\textsubscript{4} & 1282.343 &
-0.12186 & -0.07417 \\
(\textsuperscript{4}F)
4s~\textsuperscript{5}F\textsubscript{3}-(\textsuperscript{4}G) 4p
\textsuperscript{5}G\textsuperscript{o}\textsubscript{4} & 1280.534 &
-0.32532 & 0.27741 \\
(\textsuperscript{4}G)
4s~\textsuperscript{5}G\textsubscript{4}-(\textsuperscript{4}G) 4p
\textsuperscript{5}H\textsuperscript{o}\textsubscript{5} & 1281.070 &
-0.16038 & -0.2337 \\
(\textsuperscript{4}P)
4s~\textsuperscript{5}P\textsubscript{1}-(\textsuperscript{4}P) 4p
\textsuperscript{5}P\textsuperscript{o}\textsubscript{2} & 1280.561 &
-0.09343 & -0.23370 \\
(\textsuperscript{2}I)
4s~\textsuperscript{3}I\textsubscript{6}-(\textsuperscript{2}I) 4p
\textsuperscript{3}I\textsuperscript{o}\textsubscript{7} & 1281.734 &
0.07837 & 0.27157 \\
(\textsuperscript{2}I)
4s~\textsuperscript{3}I\textsubscript{7}-(\textsuperscript{2}I) 4p
\textsuperscript{3}I\textsuperscript{o}\textsubscript{7} & 1283.374 &
-0.16510 & -0.32365 \\
(\textsuperscript{2}G1) 4s~\textsuperscript{3}G\textsubscript{4}-( G1)
4p \textsuperscript{3}H\textsuperscript{o}\textsubscript{5} & 1283.858 &
-0.02032 & -0.09198 \\
(\textsuperscript{2}H)
4s~\textsuperscript{3}H\textsubscript{5}-(\textsuperscript{2}H) 4p
\textsuperscript{3}I\textsuperscript{o}\textsubscript{6} & 1285.979 &
-0.18160 & -0.23370 \\
(\textsuperscript{2}I)
4s~\textsuperscript{3}I\textsubscript{7}-(\textsuperscript{2}I) 4p
\textsuperscript{3}K\textsuperscript{o}\textsubscript{8} & 1288.962 &
-0.49035 & -0.23712 \\
(\textsuperscript{4}G)
4s~\textsuperscript{5}G\textsubscript{3}-(\textsuperscript{4}G) 4p
\textsuperscript{5}H\textsuperscript{o}\textsubscript{4} & 1287.258 &
0.17673 & -0.35173 \\
(\textsuperscript{4}F)
4s~\textsuperscript{5}F\textsubscript{2}-(\textsuperscript{4}P) 4p
\textsuperscript{5}P\textsuperscript{o}\textsubscript{3} & 1288.419 &
0.07798 & -0.21689 \\
(\textsuperscript{4}D)
4s~\textsuperscript{5}D\textsubscript{3}-(\textsuperscript{4}F) 4p
\textsuperscript{5}G\textsuperscript{o}\textsubscript{3} & 1289.834 &
-0.09813 & 0.27741 \\
(\textsuperscript{6}S)
4s~\textsuperscript{7}S\textsubscript{3}-(\textsuperscript{4}D) 4p
\textsuperscript{5}F\textsuperscript{o}\textsubscript{4} & 1287.252 &
0.03657 & -0.29248 \\
(\textsuperscript{4}G)
4s~\textsuperscript{5}G\textsubscript{2}-(\textsuperscript{6}S) 4p
\textsuperscript{7}P\textsuperscript{o}\textsubscript{2} & 1293.210 &
-0.12176 & 0.27157 \\
(\textsuperscript{2}I)
4s~\textsuperscript{3}I\textsubscript{6}-(\textsuperscript{4}G) 4p
\textsuperscript{5}H\textsuperscript{o}\textsubscript{3} & 1293.367 &
-0.32522 & -0.34136 \\
(\textsuperscript{2}I)
4s~\textsuperscript{3}I\textsubscript{5}-(\textsuperscript{2}I) 4p
\textsuperscript{3}I\textsuperscript{o}\textsubscript{6} & 1292.415 &
-0.16028 & -0.23712 \\
(\textsuperscript{2}I)
4s~\textsuperscript{3}I\textsubscript{7}-(\textsuperscript{2}I) 4p
\textsuperscript{3}I\textsuperscript{o}\textsubscript{6} & 1295.037 &
-0.09333 & -0.23712 \\
(\textsuperscript{2}S)
4s~\textsuperscript{3}S\textsubscript{1}-(\textsuperscript{2}I) 4p
\textsuperscript{3}I\textsuperscript{o}\textsubscript{6} & 1301.477 &
0.07847 & -0.23370 \\
(\textsuperscript{2}D3)
4s~\textsuperscript{3}D\textsubscript{3}-(\textsuperscript{2}S) 4p
\textsuperscript{3}P\textsuperscript{o}\textsubscript{2} & 1301.610 &
-0.16500 & 0.27602 \\
(\textsuperscript{4}D)
4s~\textsuperscript{5}D\textsubscript{1}-(\textsuperscript{2}D3) 4p
\textsuperscript{3}F\textsuperscript{o}\textsubscript{4} & 1300.660 &
-0.02022 & -0.15173 \\
(\textsuperscript{2}G2)
4s~\textsuperscript{3}G\textsubscript{5}-(\textsuperscript{4}D) 4p
\textsuperscript{5}F\textsuperscript{o}\textsubscript{2} & 1327.435 &
-0.18150 & -0.07966 \\
(\textsuperscript{6}S)
4s~\textsuperscript{5}S\textsubscript{2}-(\textsuperscript{2}H) 4p
\textsuperscript{3}I\textsuperscript{o}\textsubscript{6} & 1314.483 &
-0.49025 & -0.14136 \\
(\textsuperscript{2}F1)
4s~\textsuperscript{3}F\textsubscript{2}-(\textsuperscript{6}S) 4p
\textsuperscript{5}P\textsuperscript{o}\textsubscript{1} & 1321.210 &
0.17683 & -0.32365 \\
(\textsuperscript{2}I)
4s~\textsuperscript{3}I\textsubscript{7}-(\textsuperscript{2}F1) 4p
\textsuperscript{3}G\textsuperscript{o}\textsubscript{3} & 1317.314 &
0.07808 & -0.28995 \\
(\textsuperscript{4}G)
4s~\textsuperscript{5}G\textsubscript{6}-(\textsuperscript{2}I) 4p
\textsuperscript{3}K\textsuperscript{o}\textsubscript{7} & 1313.133 &
-0.09803 & -0.07417 \\
(\textsuperscript{6}S)
4s~\textsuperscript{5}S\textsubscript{2}-(\textsuperscript{4}G) 4p
\textsuperscript{5}G\textsuperscript{o}\textsubscript{6} & 1320.665 &
0.03667 & -0.35173 \\
(\textsuperscript{4}G)
4s~\textsuperscript{5}G\textsubscript{6}-(\textsuperscript{6}S) 4p
\textsuperscript{5}P\textsuperscript{o}\textsubscript{2} & 1316.544 &
-0.12166 & -0.23370 \\
(\textsuperscript{4}F)
4s~\textsuperscript{3}F\textsubscript{2}-(\textsuperscript{4}G) 4p
\textsuperscript{5}G\textsuperscript{o}\textsubscript{5} & 1324.771 &
-0.32512 & -0.29248 \\
(\textsuperscript{4}F)
4s~\textsuperscript{3}F\textsubscript{4}-(\textsuperscript{4}F) 4p
\textsuperscript{3}G\textsuperscript{o}\textsubscript{3} & 1326.158 &
-0.16018 & -0.32365 \\
(\textsuperscript{4}G)
4s~\textsuperscript{5}G\textsubscript{5}-(\textsuperscript{4}F) 4p
\textsuperscript{3}G\textsuperscript{o}\textsubscript{5} & 1316.841 &
-0.09323 & -0.07417 \\
(\textsuperscript{2}H)
4s~\textsuperscript{3}H\textsubscript{5}-(\textsuperscript{4}G) 4p
\textsuperscript{5}G\textsuperscript{o}\textsubscript{5} & 1321.164 &
0.07857 & -0.23712 \\
(\textsuperscript{2}I)
4s~\textsuperscript{3}I\textsubscript{6}-(\textsuperscript{2}H) 4p
\textsuperscript{3}H\textsuperscript{o}\textsubscript{6} & 1323.586 &
-0.16490 & -0.07966 \\
(\textsuperscript{4}G)
4s~\textsuperscript{3}G\textsubscript{3}-(\textsuperscript{2}I) 4p
\textsuperscript{3}K\textsuperscript{o}\textsubscript{6} & 1319.175 &
-0.02012 & -0.12371 \\
(\textsuperscript{4}G)
4s~\textsuperscript{5}G\textsubscript{2}-(\textsuperscript{4}G) 4p
\textsuperscript{5}G\textsuperscript{o}\textsubscript{3} & 1319.047 &
-0.18140 & -0.22365 \\
(\textsuperscript{2}D3)
4s~\textsuperscript{3}D\textsubscript{3}-(\textsuperscript{4}G) 4p
\textsuperscript{5}G\textsuperscript{o}\textsubscript{2} & 1326.593 &
-0.49015 & -0.18995 \\
(\textsuperscript{4}D)
4s~\textsuperscript{3}D\textsubscript{1}-(\textsuperscript{2}D3) 4p
\textsuperscript{3}F\textsuperscript{o}\textsubscript{3} & 1325.804 &
0.17693 & -0.23370 \\
(\textsuperscript{4}D)
4s~\textsuperscript{3}D\textsubscript{2}-(\textsuperscript{4}D) 4p
\textsuperscript{3}F\textsuperscript{o}\textsubscript{2} & 1327.800 &
0.07818 & -0.29248 \\
(\textsuperscript{2}D3)
4s~\textsuperscript{3}D\textsubscript{2}-(\textsuperscript{4}D) 4p
\textsuperscript{3}F\textsuperscript{o}\textsubscript{2} & 1326.971 &
-0.09793 & -0.34136 \\
(\textsuperscript{2}I)
4s~\textsuperscript{1}I\textsubscript{6}-(\textsuperscript{2}D3) 4p
\textsuperscript{3}F\textsuperscript{o}\textsubscript{3} & 1334.234 &
0.03677 & -0.21689 \\
(\textsuperscript{6}S)
4s~\textsuperscript{5}S\textsubscript{2}-(\textsuperscript{2}I) 4p
\textsuperscript{1}K\textsuperscript{o}\textsubscript{7} & 1330.722 &
-0.12156 & -0.23712 \\
(\textsuperscript{2}F1)
4s~\textsuperscript{3}F\textsubscript{3}-(\textsuperscript{6}S) 4p
\textsuperscript{5}P\textsuperscript{o}\textsubscript{3} & 1334.742 &
-0.32502 & -0.07417 \\
(\textsuperscript{2}H)
4s~\textsuperscript{3}H\textsubscript{6}-(\textsuperscript{2}F1) 4p
\textsuperscript{1}G\textsuperscript{o}\textsubscript{4} & 1329.470 &
-0.16008 & -0.33463 \\
(\textsuperscript{4}G)
4s~\textsuperscript{3}G\textsubscript{3}-(\textsuperscript{2}H) 4p
\textsuperscript{3}H\textsuperscript{o}\textsubscript{6} & 1339.956 &
-0.09313 & -0.09198 \\
(\textsuperscript{4}P)
4s~\textsuperscript{5}P\textsubscript{2}-(\textsuperscript{4}G) 4p
\textsuperscript{3}H\textsuperscript{o}\textsubscript{4} & 1325.669 &
0.07867 & -0.29248 \\
(\textsuperscript{4}P)
4s~\textsuperscript{5}P\textsubscript{1}-(\textsuperscript{4}P) 4p
\textsuperscript{5}D\textsuperscript{o}\textsubscript{2} & 1333.618 &
-0.16480 & -0.29248 \\
(\textsuperscript{4}G)
4s~\textsuperscript{3}G\textsubscript{4}-(\textsuperscript{4}P) 4p
\textsuperscript{5}D\textsuperscript{o}\textsubscript{1} & 1344.944 &
-0.02002 & -0.32365 \\
(\textsuperscript{2}I)
4s~\textsuperscript{1}I\textsubscript{6}-(\textsuperscript{4}G) 4p
\textsuperscript{3}H\textsuperscript{o}\textsubscript{5} & 1338.529 &
-0.18130 & -0.33463 \\
(\textsuperscript{2}H)
4s~\textsuperscript{3}H\textsubscript{6}-(\textsuperscript{2}I) 4p
\textsuperscript{1}H\textsuperscript{o}\textsubscript{5} & 1331.290 &
-0.49005 & -0.18995 \\
(\textsuperscript{4}G)
4s~\textsuperscript{3}G\textsubscript{5}-(\textsuperscript{2}H) 4p
\textsuperscript{3}G\textsuperscript{o}\textsubscript{5} & 1351.158 &
0.17703 & 0.27602 \\
(\textsuperscript{4}D)
4s~\textsuperscript{5}D\textsubscript{4}-(\textsuperscript{4}G) 4p
\textsuperscript{3}H\textsuperscript{o}\textsubscript{6} & 1351.287 &
0.07828 & -0.32365 \\
(\textsuperscript{4}G)
4s~\textsuperscript{3}G\textsubscript{5}-(\textsuperscript{4}P) 4p
\textsuperscript{5}P\textsuperscript{o}\textsubscript{3} & 1349.698 &
-0.35173 & -0.07417 \\
(\textsuperscript{4}G)
4s~\textsuperscript{3}G\textsubscript{3}-(\textsuperscript{4}G) 4p
\textsuperscript{3}F\textsuperscript{o}\textsubscript{4} & 1356.602 &
0.12814 & -0.34136 \\
(\textsuperscript{4}G)
4s~\textsuperscript{3}G\textsubscript{4}-(\textsuperscript{4}G) 4p
\textsuperscript{3}F\textsuperscript{o}\textsubscript{2} & 1344.944 &
-0.07966 & -0.19248 \\
\end{longtable}
\twocolumngrid
\textbf{Figure 1} shows the distribution of measured redshifts \(z_{\rm obs}\) as a function of the expected gravitational redshift, illustrating constraints on dimensionless fundamental constants (\(\alpha\) and \(G\)). Some discrepancies are evident among different methods for estimating systematic errors, reflecting the incomplete control of certain uncertainties and differences in analytical approaches and laboratory wavelength measurements \cite{le_2021_GRG,Clara_2020,MARTINS2019100301,Bagdonaite_2015_PRL,Seager_2003,Marel1993ANM}. In this study, we used observational spectra with fractional uncertainties of \(\sim 1 \times 10^{-6}\) and laboratory wavelengths accurate to \(1 \times 10^{-7}\). Model--data agreement was optimized via a nonlinear least-squares (NLS) fitting procedure. A Gaussian framework was employed to estimate errors on \(\alpha\), providing improved precision relative to previous studies. Given that both spectral fluxes and wavelengths are strictly positive, we carefully examined the data for deviations from a perfect Gaussian distribution. Any non-Gaussian features were accounted for in the analysis, ensuring robust and unbiased error estimates. Our results demonstrate that high-resolution white dwarf spectra provide some of the most stringent limits on potential spatial and temporal variations of $G$. Within GUT-inspired frameworks, changes in \(\alpha\) are correlated with variations in \(G\), allowing constraints on one to inform the other. By analyzing the Stark-broadened profiles of 120 ultraviolet Ni\,V lines in G191--B2B, we determine a fractional variation of $\dot{G}/G = (-0.014 \pm 0.016) \times 10^{-15}\,\mathrm{yr}^{-1}$, corresponding to a gravitational redshift of $z_{\mathrm{abs}} \approx 8.47 \times 10^{-5}$. Combined with complementary astrophysical observations and precision laboratory measurements, these findings provide a powerful framework for testing potential variations of fundamental constants. The full dataset is provided in the supplementary material (\textbf{Table~1}).  

\begin{figure}
\centering
\IfFileExists{fig.png}{%
  \includegraphics[width=0.48\textwidth]{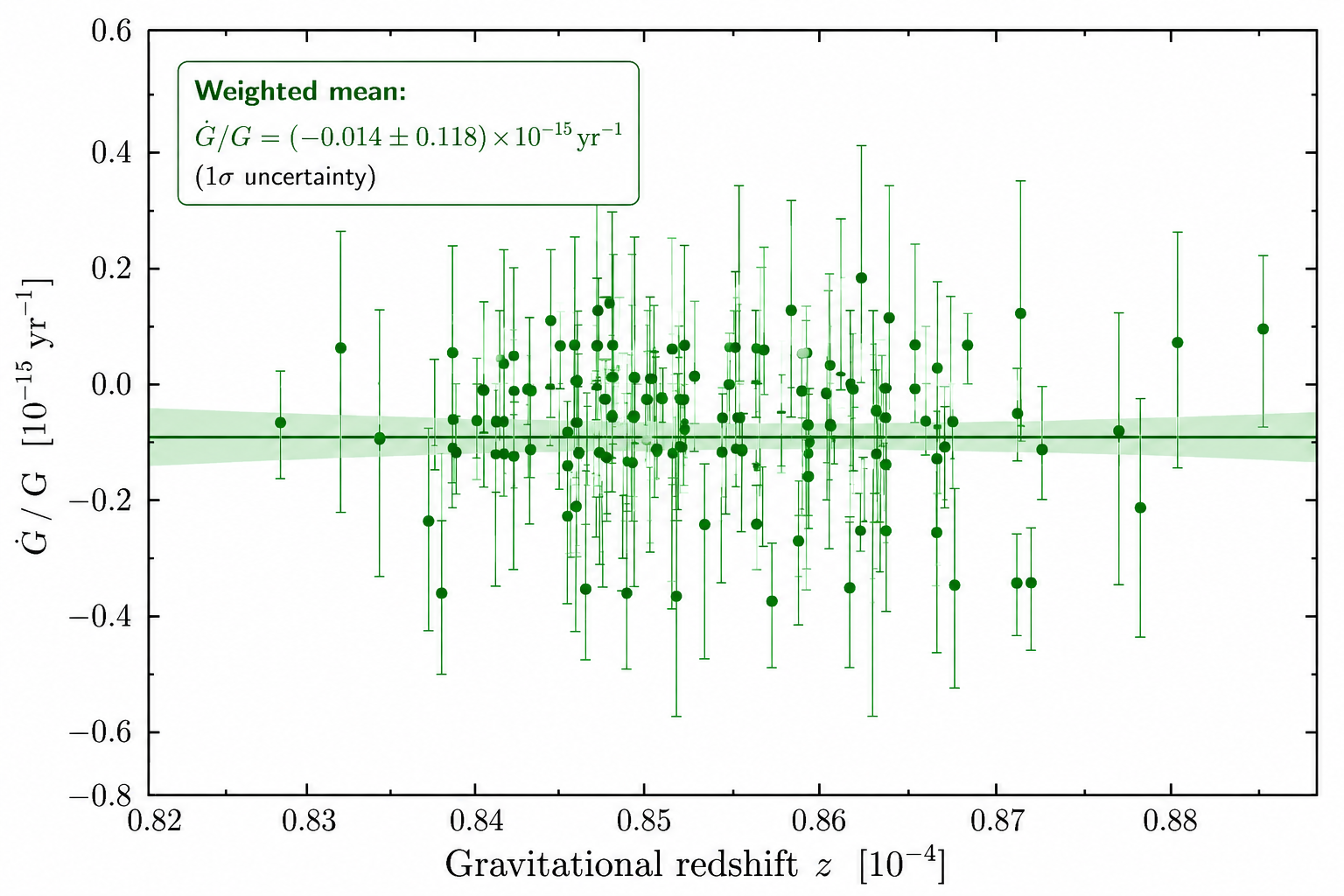}%
}{%
  \fbox{\parbox[c][4cm][c]{0.65\textwidth}{\centering Figure file \texttt{fig.png} not supplied.}}%
}
\caption{Fractional time variation of the gravitational constant, $\dot{G}/G$, as a function of gravitational redshift. Each point represents the Stark-broadened redshift measured from 120 ultraviolet Ni\,V lines in the white dwarf G191-B2B. The analysis yields a constraint of $\dot{G}/G = (-0.014 \pm 0.016) \times 10^{-15}\,\mathrm{yr}^{-1}$.}
\end{figure} 

For our analysis, the observational data consist of the measured redshifts of individual Ni\,V absorption lines in the ultraviolet spectrum of G191$-$B2B. Each transition $i$ provides an observed redshift $z_i$ and an associated $1\sigma$ uncertainty $\sigma_i$, as listed in \textbf{Table~1}. These observed redshifts include the white-dwarf gravitational redshift, possible kinematic contributions, and any additional offset that may arise if the fundamental constants vary. Variations in the fine-structure constant are linked to a variation of the gravitational constant through $\Delta G / G = \left[ \frac{8}{5}R + \frac{2}{5}(1+S) \right] \Delta\alpha / \alpha$, where $R$ and $S$ are phenomenological parameters motivated by GUTs; for given $R$, $S$, and $\Delta\alpha / \alpha$, the corresponding fractional variation in $G$ is obtained, and constraining $\Delta G / G$ is the main goal of this work. To statistically compare the observational data with theoretical expectations, we compute the predicted redshift $z_{{\rm model},i}$ for each Ni\,V transition $i$, given a set of model parameters, and the difference between $z_i$ and $z_{{\rm model},i}$ quantifies how well the parameters explain the data. We assume that the measurement uncertainties are independent and Gaussian distributed, so the probability of observing $z_i$ given the model $z_{{\rm model},i}$ is $p(z_i \,|\, z_{{\rm model},i}, \sigma_i) = \frac{1}{\sqrt{2 \pi \sigma_i^2}} \exp\big[- (z_i - z_{{\rm model},i})^2 / (2 \sigma_i^2) \big]$, and the joint likelihood of all lines is the product of individual probabilities. Taking the logarithm of the likelihood gives $\ln \mathcal{L} = -\frac{1}{2} \sum_i (z_i - z_{{\rm model},i})^2 / \sigma_i^2 - \sum_i \frac{1}{2} \ln (2 \pi \sigma_i^2)$; in practice, the constant term is often ignored, so $\ln \mathcal{L} \approx -\frac{1}{2} \sum_i \left[ (z_i - z_{{\rm model},i}) / \sigma_i \right]^2$, which is equivalent to minimizing the usual $\chi^2 = \sum_i (z_i - z_{{\rm model},i})^2 / \sigma_i^2$. A Bayesian framework is adopted to infer the posterior probability distribution of the parameters, applying Gaussian priors to $R$ and $S$ (mean values $R=273$, $S=630$, standard deviations 86 and 230) and a broad non-informative prior for $\Delta\alpha / \alpha$. From the resulting posterior distribution, the allowed range of $\Delta G / G$ is obtained, and assuming a linear temporal evolution over the characteristic age of the universe, this constraint is converted into a limit on $\dot{G}/G$.

\section {Results and Discussions}
We present the constraints obtained from the Ni\,V redshift
measurements using the statistical framework described. The
resulting probability distributions for $R$, $S$, $\Delta\alpha/\alpha$, and the derived quantity $\Delta G / G$ are obtained from the Bayesian analysis, and these are used to infer a corresponding limit on the temporal variation of the gravitational constant. Applying the methodology described, we obtain a constraint on the temporal variation of Newton's gravitational constant: $\dot{G}/G = (-0.014 \pm 0.016) \times 10^{-15}~\text{yr}^{-1}$, which is consistent with a stringent variation at the current level of sensitivity. This measurement probes a strong-gravity environment characterized by a dimensionless gravitational potential of order $\phi \sim 10^{4}$, far exceeding that accessible in Solar-System experiments. The Ni\,V spectral lines in the white dwarf G191-B2B, analyzed with fractional uncertainties of $1\times10^{-6}$ (spectra) and $1\times10^{-7}$ (laboratory wavelengths), were used to extract this constraint. In the fitting procedure, key parameters such as \((N, z_{\text{abs}}, b = \sqrt{2} \sigma)\) were utilized, assuming that the components correspond to the same transitions for all Lyman transitions of Ni V. The redshift scale or velocity was applied to determine the positions of Ni V lines, and $\dot{G}/G$ was treated as a fitting parameter.
The inference on model parameters is performed within a Bayesian framework. The likelihood is constructed assuming Gaussian uncertainties on the observed line redshifts, with the theoretical model providing predicted redshifts as functions of the parameters \(\Delta\alpha/\alpha\), \(R\), and \(S\). Uniform priors are adopted for all free parameters over broad ranges that encompass values allowed by current theoretical and experimental constraints. The joint posterior distribution is sampled using a Markov Chain Monte Carlo algorithm, from which credible intervals on combinations of parameters and derived quantities, such as \(\Delta G/G\), are obtained. This Bayesian approach makes explicit the assumptions on priors and the treatment of uncertainties. We also considered whether stellar atmospheric effects in G191-B2B could
contribute to the observed dispersion among individual Ni V line redshifts.
In particular, radiative levitation, vertical abundance stratification, and
pressure variations may cause different transitions to form at slightly
different depths in the atmosphere, where the local gravitational potential
may differ. We examined whether the residuals between the measured and
model redshifts show any trend with the line properties, but no statistically
significant correlation was identified. This suggests that atmospheric
structure is unlikely to dominate the observed scatter, although a small
contribution cannot be completely excluded. We emphasize that the parameters \(R\), \(S\), and \(\Delta\alpha/\alpha\) are not independently constrained by the redshift measurements alone due to intrinsic degeneracies in how these parameters affect the predicted line shifts. Specifically, different combinations of \(R\) and \(S\) can produce similar effects on the inferred \(\Delta G/G\) for a given \(\Delta\alpha/\alpha\), complicating independent determination. To reduce this, we adopt physically motivated priors on \(R\) and \(S\) based on representative grand-unified scenarios, and we focus our conclusions on the derived constraint on \(\Delta G/G\), which is sensitive to these degeneracies given the adopted priors. A similar strategy of combining sensitivity coefficients with priors has been used in related studies of varying constants in astrophysical settings. Ni\,V lines are especially valuable for this study because of their prevalence in white dwarf spectra and their high sensitivity to potential variations in fundamental constants, including the gravitational constant $G$. The small separations between the Ni V lines help minimize systematic effects, thereby enhancing the precision in detecting potential spatial or temporal changes in \(G\) with high accuracy.

Theoretical frameworks in fundamental physics suggest that the values of fundamental constants could vary over both cosmic time and space. One possibility involves light scalar fields, which are theorized to induce such fluctuations, with their effects potentially modulated by local gravitational fields. Recent research has investigated these phenomena, focusing on possible connections between key fundamental constants, such as the fine-structure constant (\(\alpha\)), the proton-to-electron mass ratio (\(\mu\)), and the gravitational constant (\(G\)). These studies often focus on environments with intense gravitational fields, like the photospheres of white dwarf stars \cite{Gaztanaga:2001fh,Benvenuto:2004bs,MAGUEIJO_2002}. Pulsating white dwarfs, such as G117--B15A, along with advances in white dwarf asteroseismology, have played a key role in probing potential variations in \(G\), yielding stringent constraints on its rate of change with estimates such as \(|\dot{G}/G| \leq 4.10 \times 10^{-10} \, \text{yr}^{-1}\) \cite{JAMIL2009172,Garcia-Berro:2013lea,Will:2014kxa,PhysRevD.81.064018,PhysRevLett.61.1151,Damour1991OnTO,Shapiro:2009dh}. Moreover, gravitational-wave observations and pulsar binary timing, exemplified by systems like PSR~1913+16, have provided independent limits on \(\dot{G}/G\), with reported values spanning from \(1.3 \times 10^{-10}\,\mathrm{yr}^{-1}\) to approximately \(-1.8 \times 10^{-10}\,\mathrm{yr}^{-1}\) \cite{Kaspi1994HighP,PhysRevLett.77.1432,Guenther1998TestingTC,Verbiest:2008gy}. Additionally, the effects of time- or space-varying \(G\) have been explored by comparing data obtained from six telescopes. By examining the mode spectra, stringent limits on the variation rate of the gravitational constant have been obtained, yielding \(|\dot{G}/G| < 1.6 \times 10^{-12} \, \text{yr}^{-1}\) at the two-sigma confidence level. Independent analyses consistently indicate that the temporal variation \(\dot{G}/G = (-6 \pm 42) \times 10^{-13} \, \text{yr}^{-1}\) lies well within the \(10^{-12} \, \text{yr}^{-1}\) bound, supporting the robustness of these findings \cite{Le:2024kql,Salumbides_2012}.

A complementary approach to determine the gravitational redshift relies on measuring the displacement of absorption lines in the photospheric spectrum of white dwarfs, which can be inferred from the stellar mass-to-radius ratio. However, accurately extracting the gravitational redshift is challenging because Doppler shifts arising from random stellar motions along the line of sight can contaminate the measurements. To achieve this separation, astronomers use co-moving companions as comparative references, allowing them to isolate gravitational redshift from Doppler shifts and thereby place constraints on the kinematic properties---particularly the radial velocity---of white dwarfs. By averaging the gravitational redshift across a sample of field white dwarfs and accounting for Doppler shifts due to random stellar motions, it becomes possible to derive relationships between white dwarf mass, radius, and surface temperature. The spectra of white dwarfs, encompassing photospheric absorption features, atmospheric pressure conditions, and effective temperatures, provide essential diagnostics for this analysis. For example, the spectrum of the white dwarf G191-B2B, when fitted to observational data, provides useful estimates of its physical conditions. The analysis also includes Ni\,V spectral lines, which are highly sensitive to potential variations in fundamental constants, including \(G\). Overall, these results demonstrate that high-resolution white dwarf spectra offer some of the most stringent limits on possible spatial and temporal variations of fundamental constants. By carefully accounting for observational uncertainties, Doppler contributions, and degeneracies among model parameters, our analysis provides a robust framework for testing theories that predict correlated variations in \(\alpha\), \(\mu\), and \(G\). These findings complement existing astrophysical and laboratory constraints, offering new insights into the behavior of fundamental constants in strong-gravity environments.

\section {Conclusions}

In this study, we present the most precise constraints to date on the temporal variation of the gravitational constant $G$ in a strong-gravity environment, using high-resolution ultraviolet spectra of the white dwarf G191-B2B. This star possesses a surface gravity approximately $10^{4}$ times stronger than that of the Earth, providing an ideal laboratory to probe extreme gravitational fields. By analyzing 120 Ni\,V absorption lines with state-of-the-art laboratory wavelength measurements and advanced spectral modeling, we obtain
$\dot{G}/G = (-0.014 \pm 0.016) \times 10^{-15}~\text{yr}^{-1}$,
corresponding to a gravitational potential of $\phi \sim 10^{4}$ and an average redshift of $z \approx 8.47 \times 10^{-5}$. This constraint surpasses previous limits derived from pulsar timing, lunar laser ranging, Big Bang nucleosynthesis, and stellar evolution, establishing a new context for testing the constancy of $G$ under extreme gravitational conditions.

Additionally, we link potential variations in $G$ to changes in the fine-structure constant $\alpha$ through the phenomenological parameters $R$ and $S$, situating our results within theoretical frameworks such as scalar-field interactions, extra-dimensional models, and GUTs. By combining high-precision observational data, rigorous modeling, and a carefully selected high-gravity target, we minimize systematic uncertainties and deliver the most stringent limit to date on $\dot{G}/G$. These findings not only support the apparent constancy of Newton’s gravitational constant but also impose powerful constraints on models of new physics beyond the SM, including those predicting correlated variations of fundamental constants \cite{Bellinger:2019lnl,Le:2024kql,Fritzsch:2008bj,Le:2024kql,Hamdi2022FeV,Le:2025pfo,Le2025IJMPA,PhysRevLett.77.1432,PhysRevD.81.064018}.

Overall, our results demonstrate that white dwarf spectroscopy in strong-gravity regimes provides an unparalleled probe of fundamental physics, offering complementary and highly competitive tests of gravitational theories and the stability of fundamental constants over cosmological time scales \cite{Le2026H2Lyman, Le2026MuStrongGravity}. This work establishes a robust framework for future studies aiming to detect or constrain deviations from standard gravity in extreme astrophysical environments.

\section{Declaration of Competing Interest}
The authors declare that they have no conflict of interest.

\section{Funding}
This research received no external funding.

\section{Data Availability}
All data generated or analyzed during this study are included in this published article \cite{Hamdi2024NiV}.

\bibliographystyle{apsrev4-2}
\bibliography{sn-bibliography}

\end{document}